\documentclass[letter]{aa}
\usepackage{graphicx}
\usepackage{natbib} \bibpunct{(}{)}{;}{a}{}{,}
\usepackage{color}
\usepackage[varg]{txfonts}
\usepackage{rotating}
 
\def\changed{}
\def\changedA{}
\def\changedAA{}
\def\changedAAA{}
 
  \def\HII{\ion{H}{ii}} 
 \def\HeII{\ion{He}{ii}}  
   \def\CIV{\ion{C}{iv}}
    
  \def\NIV{\ion{N}{iv}} \def\NV{\ion{N}{v}}

  \newcommand{\msunpyr}{M_\odot\,\mbox{yr}^{-1}}

\newcommand{\kms}{\ifmmode{\,\mbox{km}\,\mbox{s}^{-1}}\else{km/s}\fi}
\newcommand{\msun}{\ifmmode M_{\odot} \else M$_{\odot}$\fi}
\newcommand{\rsun}{\ifmmode R_{\odot} \else R$_{\odot}$\fi}
\newcommand{\lsun}{\ifmmode L_{\odot} \else L$_{\odot}$\fi}
\newcommand{\zsun}{\ifmmode Z_{\odot} \else $Z_{\odot}$\fi}
\newcommand{\velo}{\ifmmode\varv\else$\varv$\fi}
\newcommand{\vinf}{\ifmmode\velo_\infty\else$\velo_\infty$\fi}
\newcommand{\LHE}{\ifmmode L_\ion{He}{ii} \else $L_\ion{He}{ii}$\fi}
\newcommand{\LCIV}{\ifmmode L_\ion{C}{iv} \else $L_\ion{C}{iv}$\fi}
\newcommand{\LWN}{\ifmmode L_{\rm WN5h} \else $L_{\rm WN5h}$\fi}
\newcommand{\NWN}{\ifmmode N_{\rm WN5h} \else $N_{\rm WN5h}$\fi}
\newcommand{\LWNZ}{\ifmmode L_{{\rm WNh, low} Z} \else $L_{{\rm WNh, low} Z}$\fi}
\newcommand{\NWNZ}{\ifmmode N_{{\rm WNh, low} Z} \else $N_{{\rm WNh, low} Z}$\fi}
\begin{document} 

\title{Narrow He\,II emission in star-forming galaxies at low
  metallicity}
\subtitle{Stellar wind emission from a population of very massive stars}
 
 
\author{G.\ Gr\"{a}fener\inst{} \and J.S.\  Vink\inst{}}

\institute{Armagh Observatory, College Hill, Armagh, BT61\,9DG, United
  Kingdom\label{inst1}}

\date{Received ; Accepted}
 

\abstract{In a recent study, star-forming galaxies with
  \HeII\,$\lambda$1640 emission at moderate redshifts between 2 and
  4.6 have been found to occur in two modes that are distinguished by the
  width of their \HeII\ emission lines.  Broad \HeII\ emission has
  been attributed to stellar emission from a population of evolved
  Wolf-Rayet (WR) stars.  The origin of narrow \HeII\ emission is less
  clear but has been attributed to nebular emission excited by a
  population of very hot Pop\,III stars formed in pockets of pristine
  gas at moderate redshifts.}{We propose an alternative
  scenario for the origin of the narrow \HeII\ emission, namely very
  massive stars (VMS) at low metallicity ($Z$), which form strong
  {\changedAAA but slow} WR-type stellar winds due to their proximity
  to the Eddington limit.}{We estimated the expected \HeII\ line fluxes
  {\changedAA and equivalent widths} based on wind models for VMS
  {\changedAA and Starburst99 population synthesis models} and compared
  the results with recent observations of star-forming galaxies at
  moderate redshifts.} {The observed \HeII\ line strengths {\changedAAA
    and equivalent widths} are in line with what is expected for a
  population of VMS in one or more young super-clusters located within
  these galaxies.}  {In our scenario the two observed modes of \HeII\
  emission originate from massive stellar populations in distinct
  evolutionary stages at low $Z$ ($\sim 0.01\,Z_\odot$). If this
  interpretation is correct, there is no need to postulate the
  existence of Pop\,III stars at moderate redshifts to explain the
  observed narrow \HeII\ emission. An interesting possibility is the
  existence of self-enriched VMS with similar WR-type spectra at
  extremely low $Z$.  Stellar \HeII\ emission from such very early
  generations of VMS may be detectable in future studies of
  star-forming galaxies at high redshifts {\changed with} the
  James Webb Space Telescope (JWST). The fact that the \HeII\ emission
  of VMS is largely neglected in current population synthesis models
  will generally affect the interpretation of the integrated spectra
  of young stellar populations.}

\maketitle

\section{Introduction} 
\label{intro}

A fundamental question in Astronomy is the question of the sources of
the ``First Light'' ending the Cosmic Dark Ages and beginning the
process of reionisation. The James Webb Space Telescope (JWST)
promises direct access to this critical period via observations of the
first star-forming galaxies at redshifts $z\gtrsim10$.  Dual
Ly$\alpha$ $\lambda$1216 and \HeII\ $\lambda$1640 emission is believed
to be the main indicator of the first (Pop\,III) stars formed from
pristine gas in these objects \citep[e.g.][]{joh1:09}. The reason is
that only at extremely low metallicities ($Z\lesssim10^{-5}$), young
massive stars are believed to be hot enough to excite \HeII\ in their
surrounding \HII\ regions \citep[e.g.][]{sch1:02,sch2:03}. The
investigation of star-forming galaxies with \HeII\ emission at
moderate redshifts, which are accessible with current ground-based
instrumentation, is thus an important preparation for future studies
of the first star-forming galaxies with the JWST.

Such a study has recently been performed by \citet{cas1:13}, who
identified different groups of \HeII\ emitters with narrow and broad
emission lines in a sample of star-forming galaxies at redshifts
$2<z<4.6$. While the broad emission is usually attributed to stellar
emission from Wolf-Rayet (WR) stars, \citeauthor{cas1:13} interpreted
the narrow emission as being due to {\changedAAA ionising radiation from}
Pop\,III stars that must have formed in pockets of residual pristine
gas at relatively low redshift, {\changedAAA or a stellar population that
  is rare at $z \sim 0$ but more common at $z \sim 3$}.

In this letter we discuss the possibility that the narrow \HeII\
emission might {\changedAA also be} of stellar origin. To this purpose we
discuss in Sect.\,\ref{WRZ} how the spectral properties of WR-type
emission line stars are expected to change at low $Z$, {\changedAAA
  estimate the expected line shapes and fluxes in
  Sect.\,\ref{sec:EXPECT}, and compare our results with the
  observations of \citeauthor{cas1:13} in Sect.\,\ref{sec:CASSATA}.}
Conclusions are drawn in Sect.\,\ref{CONCL}.

\section{\HeII\ emission from WR stars at different metallicities $Z$}
\label{WRZ}

The emission-line spectra of WR stars are formed in their strong,
optically thick stellar winds. As a result of the large optical depth of these
winds, ions like He\,{\sc iii} recombine already within the wind,
leading to recombination cascades with strong emission lines between
excited levels of the subordinate ions. Prominent examples are
He\,{\sc ii}\,$\lambda$1640 and \CIV\,$\lambda$1550 in the UV, and
He\,{\sc ii}\,$\lambda$4686 and \CIV\,$\lambda$5808 in the optical
wavelength range. The width of the emission lines is mainly determined
by the terminal wind speed $\varv_\infty$ of the stellar wind. To
understand why we naturally expect two populations of WR stars with
broad {\em \textup{and}} narrow emission lines at low $Z,$ we need to discuss
the origin of the different types of WR-type stellar winds.

\subsection{WR populations in the Galaxy and LMC}
\label{sec:wrgal}

Comprehensive studies of the local population of WR stars of the
nitrogen sequence (WN stars) have been performed for the Galaxy by
\citet{ham1:06} and for the Large Magellanic Cloud (LMC) by
\citet{hai1:14}. Both works identified two different groups of WN
stars. The first group are classical WN stars with H-poor/H-free
surface compositions and hot temperatures up to 140\,kK, in agreement
with a core He-burning evolutionary stage. The second group are the
so-called WNh stars with relatively H-rich surface compositions and
cooler temperatures, in agreement with a core H-burning evolutionary
stage. The WNh stars are further characterised by remarkably high
luminosities in excess of $\sim 10^{5.9}L_\odot$, while the group of
classical WN stars has luminosities below this value.

The dividing luminosity at roughly $10^{5.9}L_\odot$ suggests that the
classical WN stars are core He-burning objects in a post-red
supergiant (post-RSG) phase.  This picture is further supported by the
work of \citet{san1:12}, who found that the WR stars of the carbon
sequence (WC stars), which are showing the products of He-burning at
their surface, form a natural succession of the sequence of classical
WN stars in the HR diagram.

In this work we are mainly interested in the group of WNh stars with
luminosities in the range $10^{5.9}L_\odot \lesssim L \lesssim
10^{6.9}L_\odot$ \citep[cf.][]{ham1:06,cro1:10,hai1:14,bes1:14}. The
HRD positions of these objects suggest that they are very massive
stars (VMS) with initial masses in excess of $\sim 100\,M_\odot$, and
reaching up to $300\,M_\odot$ \citep[cf.][]{vin1:15}.  In their LMC
sample, \citet{hai1:14} found 12\,\% of all putatively single WN stars
in the WNh stage. Combined with their high luminosities, this suggests
that WNh stars have substantial impact on the integrated \HeII\
emission of star-forming galaxies.

\subsection{WR mass loss at low metallicities $Z$}
\label{sec:WRZ}

Observational evidence for a $Z$-dependence of WR mass-loss has been
found by comparisons of the sample of WC stars with known distances in
the Galaxy and LMC by \citet{gra1:98} and \citet{cro1:02}. In the
following, we invoke the results of theoretical studies to understand
how the properties of WR stars are expected to change over a broad
range of $Z$.

The winds of classical WR stars have been modelled by \citet{gra1:05}
(for early WC subtypes) and \citet{vin1:05} (for late WC and WN
subtypes). Although there are still problems to reproduce the winds of
intermediate spectral subtypes \citep[cf.][]{gra1:05}, these works
suggest that the winds of WR stars are radiatively driven,
predominantly by millions of spectral lines of the iron-group
elements. As a consequence, the mass-loss rates of WR stars are
expected to depend on the environment metallicity $Z$ \citep{vin1:05}.

The winds of luminous WNh stars have been modelled by \citet{gra1:08}
and \citet{vin1:11}. These works suggest that the proximity to the
Eddington limit plays a key role for the formation of WR-type stellar
winds. This has been further supported in systematic studies of
samples of VMS in young clusters by \citet{gra1:11} and
\citet{bes1:14}.  According to \citet{gra1:08}, Eddington factors of
the order of $\Gamma_{\rm e}\approx 0.5$ are required to initiate
WR-type stellar winds at solar metallicity $Z_\odot$ (here
$\Gamma_{\rm e}=(\chi_{\rm e} L)/(4\pi c G M)$ denotes the
classical Eddington factor that is defined with respect to the
opacity of free electrons $\chi_{\rm e}$).  The required Eddington
factors suggest that the masses of WNh stars are higher than those of
core He-burning stars, meaning that WNh stars are indeed VMS in the
phase of core H-burning.

How does the spectral appearance of the two groups of WR stars change
at low metallicities?  As a result of their dependence on the iron-group
opacities, the winds of classical WR stars are expected to be weaker at
low $Z$ \citep{vin1:05}, resulting in significantly weaker emission
lines \citep{cro1:06}. However, the situation is slightly
different for WC stars because their surfaces are substantially self-enriched with
the products of He-burning, namely C, O, and most likely Ne and Mg.
According to \citet{vin1:05}, these elements gain in importance for
$Z\lesssim 0.1\,Z_\odot$, and may even dominate WR wind driving for $Z
\lesssim 10^{-3}\,Z_\odot$. WC stars may thus be the main source of
broad \HeII\ emission from classical WR stars at low $Z$. Because
of the
high carbon abundances of WC stars, this emission will most likely be
accompanied by strong C\,{\sc iv} emission.

For the group of very massive WNh stars, the dependence on the
environment metallicity $Z$ has been discussed by \citet{gra1:08}.
According to their models, the proximity to the Eddington limit is
mandatory to support WR-type winds in this regime. The required
Eddington factors increase for decreasing metallicity, from
$\Gamma_{\rm e}\approx 0.5$ for $Z=Z_\odot$, to $\Gamma_{\rm e}\approx
0.85$ for $Z=0.01 Z_\odot$. As a consequence, the resulting effective
escape velocities decrease, and very low terminal wind velocities of
only few 100\,km/s are predicted for WNh stars in this regime. At the
same time, the mass-loss rates do not change significantly, resulting
in significantly higher wind densities and much stronger and narrower
WR emission lines than for classical WR stars. {\changedAAA Very massive
  WNh stars at low $Z$} may thus be the sources of the narrow \HeII\
emission observed by \citet{cas1:13}. In contrast to classical WC
stars, they have very low carbon abundances and are not expected to
show detectable \CIV\ emission.

\section{Expected \HeII\ emission from populations of VMS}
\label{sec:EXPECT}

{\changedAAA Because of their short lifetimes (2-3\,Myr), VMS are
  predominantly found in young massive clusters. To estimate the line
  fluxes expected from populations of VMS at cosmological distances, we
  start with an examination of young massive clusters in our
  neighbourhood and their stellar \HeII\ emission in
  Sect.\,\ref{sec:LOCAL} and extrapolate our results to low $Z$ in
  Sect.\,\ref{sec:LOWZ}.}

\subsection{VMS in young massive clusters}
\label{sec:LOCAL}

\citet{wof1:14} recently found evidence for \HeII\ emission from VMS
in the nearby starburst galaxy NGC\,3125 with roughly LMC metallicity.
For the young cluster A1 {\changedAA they had to invoke a population
  of VMS to explain the large observed \HeII\ equivalent width
  \citep[EW(\HeII) $\sim 7$\AA, cf.\ also][]{cha1:04} as the cluster
  is too young (3--4\,Myr) to host evolved WR stars. {\changed Based
    on different UV extinction corrections,} they determined an
  intrinsic \HeII\ line luminosity in the range
  $\log(\LHE)=39.1\,...\,40.0$ {\changedA (in erg/s)} and a cluster
  mass of $\log(M_{\rm cl}/M_\odot)= 5.1\,...\,5.9$.}

{\changedAA To investigate whether VMS can account for the large
  EW(\HeII), we also investigated R136 in the centre of 30\,Dor of
  the LMC, the most prominent starburst-like cluster in our
  neighbourhood that is spatially resolved.  Using the integrated IUE
  spectrum of the central 5\,pc from \citet{vac1:95}, we measured a
  similarly high EW(\HeII) $\sim 5$\AA\ for this object.

  {\changedA For the sum of the seven known WNh stars in this region
    \citep[five\,WN5h, one\,WN6(h), and one\,WN9ha; cf.][]{cro1:10}, we
    estimate in total $\log(\LHE) \approx 38.6$. This estimate is
    based on a synthetic model for the LMC WN5h star VFTS\,682 from
    \citet{bes1:11} with $\log(L/L_\odot)=6.5$ and $\log(\LHE)= 37.5$,
    which was scaled to the total luminosity of the seven objects 
($\log(L/L_\odot)=7.58\pm0.03$, as obtained from \citealt{cro1:10},
\citealt{hai1:14}, and \citealt{bes1:14}).}  We note that this value
is uncertain as for some of the stars different luminosities have been
obtained in different works, and one of the stars, Mk\,34, is a known
binary.

  To estimate the intrinsic continuum flux, we used Starburst99 models
  \citep{lei1:99} that are similar to those of \citet[][cf.\ their
  Sect.\,3.5]{wof1:14} but for $Z=0.008$. For a cluster mass of
  $\log(M_{\rm cl}/M_\odot) \approx 4.7$
  \citep[][]{dor1:13}\footnote{\changedA \citeauthor[][]{dor1:13} used
    the same \citet{kro1:01} IMF as in this work.}  and an age of
  2--4\,Myr, we obtain a continuum flux $\log(L_{\rm UV}) \approx
  37.74$ {\changedA (in erg\,s$^{-1}$\,\AA$^{-1}$)}.  Together with our
  estimate of $\log(\LHE) \approx 38.6,$ this results in
  EW(\HeII)\,$\approx$\,7\,\AA, in good agreement with the
  observations of R\,136 and NGC\,3125-A1. We note that the intrinsic
  continuum flux estimated by \citet{vac1:95} is even lower
  ($\log(L_{\rm UV})\approx 37.33$), which may be attributed to
  uncertainties in the UV extinction and the precise pointing of their
  observations.  }

{\changedAA We conclude that the \HeII\ emission of very massive WNh
  stars at $Z \approx Z_{\rm LMC}$ is probably strong enough to explain
  the observed strength of the \HeII\ emission of NGC\,3125-A1 and
  R136.  In the following, we use the approximate average luminosity of
  the seven WNh stars in R\,136 to define a representative luminosity
  of $\log(\LWN/L_\odot)=6.7$. With the numbers above, we then obtain
\begin{equation}
\label{eq:NWN}
\log(\LHE/({\rm erg}/{\rm s}))=\log(\NWN)+37.7,\end{equation} 
where \NWN\ denotes the equivalent number of WN5h stars.}

{\changedAA The observed \HeII\ line widths (FWHM) lie in the range of
  1000-1400\,km/s for NGC\,3125-A1 \citep{cha1:04,wof1:14} and
  $\sim$\,$2500$\,km/s for R\,136.  The individual WN5h stars in
  R\,136 have a similar spectral morphology as VFTS\,682, for which we
  estimate FWHM\,$\sim$\,2000\,km/s.  The variations in the observed
  FWHM may indicate that the real stellar populations are more
  heterogeneous than we assumed in our simplified approach.}

\subsection{\HeII\ emission at low $Z$}
\label{sec:LOWZ}

\begin{figure}[tbp]
  \parbox[b]{0.49\textwidth}{\center{\includegraphics[scale=0.43]{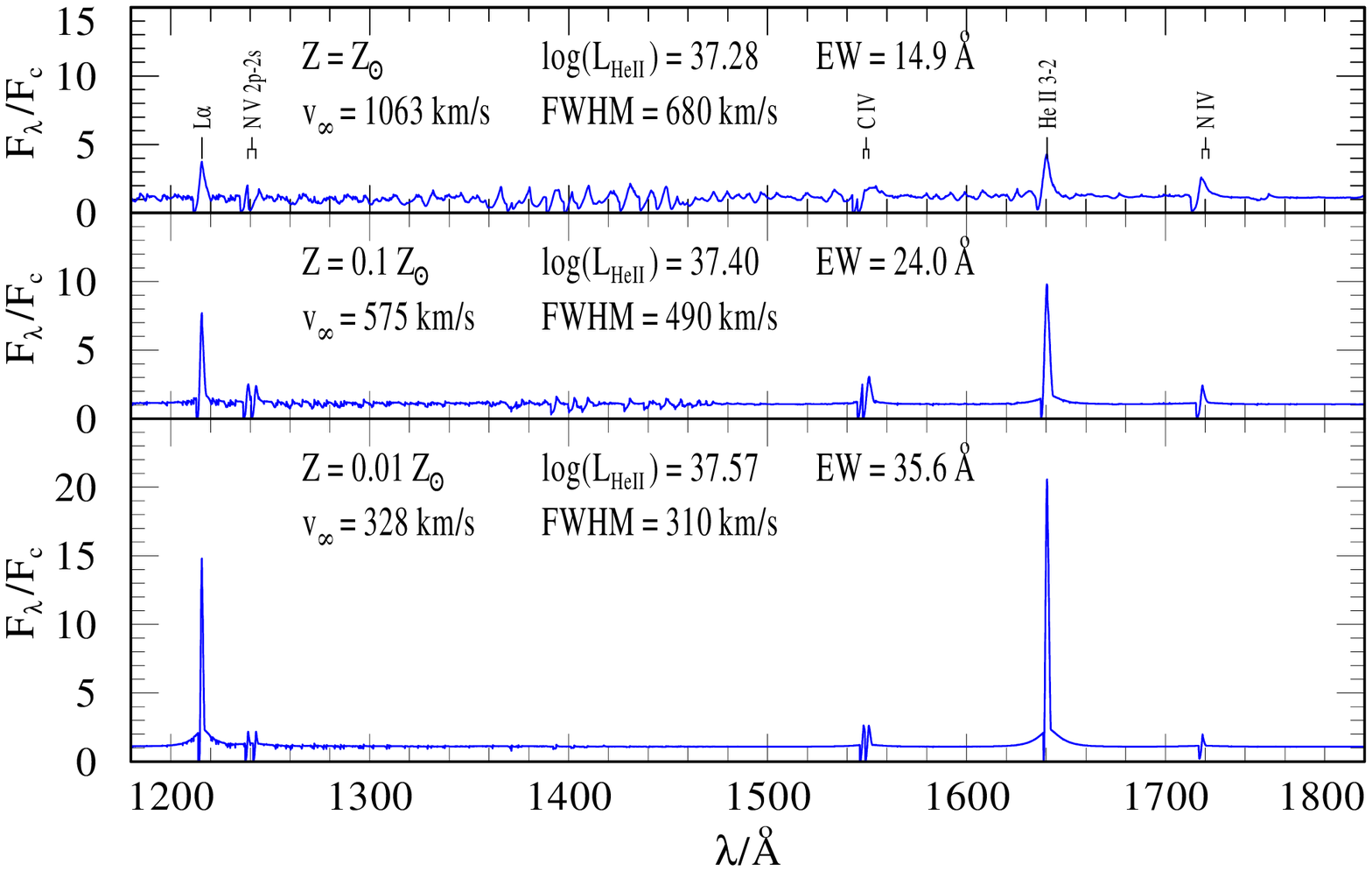}}}
  \caption{Synthetic UV spectra for different metallicities $Z$ from
    wind models for WNh stars from \citet{gra1:08}. The presented
    models are computed for a luminosity of $\log(L/L_\odot)=6.3$, a
    stellar temperature $T_\star=44.7$\,kK, and have very similar
    mass-loss rates ($\sim 1.8 \times 10^{-5}\,\msunpyr$). The given
    \HeII\ line luminosities have only been determined for the narrow-emission components of the \HeII\ line profiles. {\changedAAA
      Measurement errors for the given $\LHE$, FWHM, and EW probably
      amount to $\sim \pm 0.05$\,dex} (or $\pm10\%$).}
  \label{fig:wnlz}
\end{figure}

To estimate $\LHE$ for WNh stars at low $Z,$ we {\changedAAA computed
  synthetic UV spectra using the wind models of \citet{gra1:08}.
  {\changedAA To obtain a more realistic FWHM,} we used a line broadening
  velocity of 10\,km/s instead of the {\changedAA artificially
    enhanced} value of 100\,km/s in the original work.} In
Fig.\,\ref{fig:wnlz} we show three models with $Z$\,$=$\,$Z_\odot$,
$0.1\,Z_\odot$, and $0.01\,Z_\odot$ \citep[cf.\ model grid 2 in
Table\,2 of][]{gra1:08}.  For these models, $\Gamma_{\rm e}$ has been
adjusted to reproduce very similar mass-loss rates for all three
metallicities\footnote{As the luminosities and surface abundances are
  kept fixed, this change in $\Gamma_{\rm e}$ translates into
  different stellar masses $M$ for each object.}.  As discussed in
Sect.\,\ref{sec:WRZ}, the models with low $Z$ have lower terminal wind
velocities $\varv_\infty$ and thus higher wind densities. This leads
to stronger and narrower \HeII\ emission at lower $Z$ (cf.\
Fig.\,\ref{fig:wnlz}). {\changedAAA The resulting line fluxes and FWHM
  are indicated in Fig.\,\ref{fig:wnlz}. We note that the wind models
  seem to underestimate $\varv_\infty$ at $Z=Z_\odot$ \citep[cf.\ the
  discussion in][]{gra1:08}. The important point is thus the predicted
  {\em \textup{qualitative}} behaviour, that is, the strong decrease of
  $\varv_\infty$ vs.\ $Z$.}

The \HeII\ line profiles in Fig.\,\ref{fig:wnlz} consist of a narrow
emission component, a comparatively weak blue-shifted P-Cygni
absorption, and broad line wings that are caused by multiple electron
scattering of line photons in the dense ionised wind. In addition to
the iron forest between 1300 and 1500\,\AA, which is most prominent
at high $Z$, the spectra show \CIV, \NIV, and \NV\ features whose
strength decreases for lower $Z$. We note that the models have been
computed for an N-enriched surface composition, which will affect the
relative strength of the \CIV\ vs.\ \NIV\ and \NV\ features.
Furthermore, the resulting \HeII\ line fluxes will depend on the
adopted stellar temperature $T_\star$.

For low metallicities $Z=0.1$ and 0.01, we find \HeII\ line
luminosities of the order of $\log(\LHE)\approx 37.5$ {\changedAAA for
  a given stellar luminosity of $\log(L/L_\odot=6.3)$.  {\changedAA
    Adopting the same representative stellar luminosity as in
    Eq.\,\ref{eq:NWN}, we obtain
\begin{equation}
\label{eq:NWNZ}
  \log(\LHE/({\rm erg}/{\rm s}))=\log(\NWNZ)+37.9,
\end{equation}
that is, a slightly higher value than the one that we obtained
for LMC metallicity.}

While $\LHE$} increases for lower $Z,$ the strength of the observable
metal lines decreases.  Most notably, all metal lines are
substantially weaker than \HeII\, and will not be detectable in the
integrated spectra of \HeII\ emitting galaxies, as observed by
\citet{cas1:13}.

{\changedAAA We conclude that WNh stars at low $Z$ can produce narrow
  \HeII\ emission with similar or slightly stronger line fluxes than
  those discussed in Sect.\,\ref{sec:LOCAL} if the high required
  Eddington factors $\Gamma_{\rm e}$ can be maintained.
  Theoretically, high {\em \textup{effective}} $\Gamma_{\rm e}$ are indeed
  expected due to increased stellar rotation rates at low $Z$
  \citep[e.g.,][]{mey1:02}.  Adopting similar relative numbers and
  luminosities of WNh stars as in Sect.\,\ref{sec:LOCAL}, we would
  thus expect similar or slightly higher line fluxes and EWs than those discussed in Sect.\,\ref{sec:LOCAL} for young starbursts at low
  $Z$.}

{\changedAAA We also note that} rotating massive stars in extremely
metal-poor environments ($Z \sim 10^{-5}...\,10^{-8}\,Z_\odot$) are
expected to reveal primary nitrogen with mass fractions of about 1\% at
their surface \citep{yoo1:05,mey1:06}.  \citet{gra1:08} showed
that such self-enriched objects can develop very similar winds as WNh
stars at low metallicity (roughly corresponding to $Z \sim
0.02\,Z_\odot$).  We thus expect that even very early stellar
generations can contain WNh stars with similar spectral properties as
discussed here. For Pop\,III stars with zero metal content,
\citet{eks1:08} showed that qualitatively different internal mixing
properties can prevent strong self-enrichment, that is, their mass-loss
properties may be different \citep[but cf.\ also the discussion
in][]{vin1:05}.

\section{Narrow \HeII\ emission in star-forming galaxies at
  moderate redshifts}
\label{sec:CASSATA}

A systematic study of star-forming galaxies with \HeII\ emission at
moderate redshifts has been performed by \citet{cas1:13}, based on
data from the VIMOS VLT Deep Survey \citep[VVDS;][]{lef3:05}. In the
total VVDS sample of $\sim 45000$ galaxy spectra, \citeauthor{cas1:13}
found 277 star-forming galaxies with secure redshifts in the range
$2<z<4.6,$ of which 39 show \HeII\,$\lambda$1640 in emission.  They
divided this sample into four groups of 11 objects with narrow \HeII\
lines, 13 objects with broad \HeII\ lines, 12 possible \HeII\ emitters,
and 3 AGN, to distinguish between high-velocity outflows of WR\,stars,
AGN, or supernova-driven winds, and the narrow nebular emission that
is expected for Pop\,III stars.

\citeauthor{cas1:13} found that narrow and broad emitters show
qualitatively different spectra. In particular, they found that only
broad \HeII\ emitters (with $\mbox{FWHM}(\HeII) \ge 663$\,km/s) show
indications of underlying P-Cygni type C\,{\sc iv} $\lambda$1550
emission. As we discussed in Sect.\,\ref{sec:WRZ}, this emission is
most likely a signature of stellar-wind emission from classical WR
stars, predominantly originating from WC subtypes at low $Z$.
{\changedAA In the group of 11 narrow emitters, they only found two} such
cases, in line with our prediction of very weak \CIV\ emission for WNh
stars at low $Z$.  {\changedAAA Our predicted FWHM(\HeII) for such
  objects in Sect.\,\ref{sec:LOWZ} is clearly within the limit for
  narrow emitters from \citeauthor{cas1:13}, even if we take possible
  uncertainties (up to a factor of $\sim$\,2) into account.}

Excluding AGN, \citeauthor{cas1:13} found \HeII\ line luminosities
{\changedAAA in the range $\log(\LHE) \approx 40\,...\,41.5$, and EWs
  of $\approx 1\,...\,7$\AA}.  They noted that, based on the
predicted nebular \HeII\ emission for Pop\,III objects from
\citet{sch2:03}, the star-formation rates required to produce
{\changedAAA the observed} line fluxes are about two orders of
magnitude lower than {\changedAAA those determined} from the
spectral energy distribution (${\rm SFR}_{\rm sed}$).

Here we furthermore note that the majority of narrow \HeII\ emitters is
found at the bottom of the \HeII\ luminosity distribution in a very
narrow range of $\log(\LHE) \approx 40\,...\,40.3$ (cf.\ Fig.\,10 in
\citet{cas1:13}).  This low-luminosity group is additionally restricted
to the lowest redshifts of $z\approx2.0\,...\,2.5$.  This implies that
these objects are near the detection limit. {\changedAAA Only three
  objects are located outside the low-luminosity group, with
  $\log(\LHE) \approx 41.0\,...\,41.5$, and $z=2.5\,...\,4.0$.}

{\changedAA The observed line fluxes for the low-luminosity group are
  remarkably similar to our expectations for massive young
  super-clusters with $M_{\rm cl}\,\sim\,10^6\,M_\odot$ at low $Z$.}
Clusters in this mass range are near the top end of the cluster
initial mass function ($\sim\,2\,\times\,10^6\,M_\odot$), as
determined for instance\ by \citet{bas1:08} for interacting galaxies and
luminous IR galaxies.  They may thus be the most luminous \HeII\
sources of this kind.
{\changedA For comparison, the observed continuum fluxes $\log(L_{\rm
    UV}) = \log(\LHE/{\rm EW}(\HeII)) \approx 39.2\,...\,40.3$ are
  similar to or higher than the Starburst99 prediction of $\log(L_{\rm
    UV}) \approx 39.04$ for a cluster with $10^6\,M_\odot$ and an age
  of 2-4\,Myr at low $Z=0.0004$. Test computations for galaxies with
  continuous SFRs at the same metallicity suggest $L_{\rm UV} \lesssim
  40.2 + \log({\rm SFR/(\msunpyr)})$, meaning that the host galaxies could
  provide a substantial contribution to the observed continuum.}

{\changedA In our scenario, the continuum of a star-forming host galaxy
  could easily become so bright that it would prevent the detection of
  \HeII.  The target selection by \citeauthor{cas1:13} may thus be
  responsible for picking the most extreme and recent starburst events
  at low $Z$ that are located in host galaxies with continua weak
  enough to enable the detection of \HeII. The latter point places
  strong constraints on the continuum flux, which may be the reason for
  the narrow range of observed i-band magnitudes
  (23.45\,...\,24.63\,mag$_i$) compared to the wide range of derived
  ${\rm SFR}_{\rm sed} \approx\,1\,...\,10^3\,\msunpyr$ of
  \citeauthor{cas1:13} (cf.\ their Table\,2).}

\section{Conclusions}
\label{CONCL}

We discussed a new scenario for the origin of narrow
\HeII\ emission in star-forming galaxies at moderate redshifts, based
on predictions of strong WR-type stellar winds with low terminal wind
speeds for very massive stars (VMS) at low metallicities ($Z\sim
0.01\,Z_\odot$).  We estimated the expected \HeII\ emission and found
that it is in line with recent observations by \citet{cas1:13}. In our
scenario the observed emission originates from a population of VMS in
one or more young super-clusters located within a variety of
star-forming galaxies at low $Z$. {\changedAAA This scenario is in
  accordance with} the observed line fluxes {\changedAAA and EWs}, the
large observed spread of star-formation rates, and the {\changedA
  different \CIV\ profiles of the narrow and broad emitters.}

If our interpretation is correct, the narrow {\em \textup{and}} broad \HeII\
emitters identified by \citeauthor{cas1:13} both originate from
massive stellar populations at low $Z$ in distinct evolutionary
stages, and there is no need to invoke the existence of Pop\,III stars at
moderate redshifts to explain the observed narrow \HeII\ emission.
The fact that the emission of VMS is largely neglected in current
population synthesis models will generally affect the interpretation
of the integrated spectra of young stellar populations \citep[cf.\
also][]{wof1:14}.

Finally, we note that the same argumentation will hold for rotating
VMS at extremely low $Z$ ($\sim 10^{-5}...\,10^{-8}Z_\odot$), for
which similarly strong winds are predicted as a result of the self-enrichment
with primary nitrogen.  The \HeII\ emission from such very early
stellar generations of VMS may be detectable in future studies of
star-forming galaxies at high redshifts with the JWST.



\end{document}